\documentclass{jfm}

\usepackage[usenames, dvipsnames]{color}
\usepackage{ar}
\usepackage{amsmath,color,graphicx,amssymb}
\usepackage{subcaption}
\usepackage{amsbsy}
\usepackage{latexsym}
\usepackage[normalem]{ulem}
\usepackage[english]{babel}
\usepackage{ifsym}
\usepackage{bbding}
\usepackage{wasysym}
\usepackage{epsfig} 
\usepackage{epsf,psfig}
\usepackage{epstopdf}
\usepackage{multirow}
\usepackage{array}
\usepackage{float}
\usepackage[english]{babel}
\usepackage{verbatim}
\usepackage{soul}
\usepackage[toc,page]{appendix}
\usepackage{setspace}
\usepackage{threeparttable}
\usepackage{booktabs,array}
\usepackage{color}
\usepackage{mathtools}
\usepackage{siunitx}
\usepackage{setspace}
\usepackage[numbers]{natbib}
\usepackage{authblk}

\definecolor{cobalt}{rgb}{0.0, 0.28, 0.67} 
\definecolor{cadetblue}{rgb}{0.37, 0.62, 0.63} 
\definecolor{carnelian}{rgb}{0.7, 0.11, 0.11}

\def\vK{von K\'{a}rm\'{a}n }

\begin{document}

\newtheorem{lemma}{Lemma}
\newtheorem{corollary}{Corollary}

\shorttitle{Scaling Laws for Three-Dimensional Pitching Propulsors} 
\shortauthor{F. Ayancik et al.} 

\title{Scaling Laws for the Propulsive Performance of Three-Dimensional Pitching Propulsors}

\author
 {
 Fatma Ayancik\aff{1}
  \corresp{\email{faa214@lehigh.edu}},
  Qiang Zhong\aff{2}, Daniel B. Quinn\aff{2}, Aaron Brandes\aff{2}, Hilary Bart-Smith\aff{2}
  \& 
  Keith W. Moored\aff{1}
  }

\affiliation
{
\aff{1}
Department of Mechanical Engineering, Lehigh University, Bethlehem, PA 18015, USA
\aff{2}
Department of Aerospace and Mechanical Engineering, University of Virginia, Charlottesville, VA 22904, USA
}
\graphicspath{{Figures/}}
\maketitle

\begin{abstract}
Scaling laws for the thrust production and energetics of self-propelled or fixed-velocity three-dimensional rigid propulsors undergoing pitching motions are presented.  The scaling relations extend the two-dimensional scaling laws presented in \cite{moored2018inviscid} 
by accounting for the added mass of a finite-span propulsor, the downwash/upwash effects from the trailing vortex system of a propulsor, and the elliptical topology of shedding trailing-edge vortices. The novel three-dimensional scaling laws are validated with self-propelled inviscid simulations and fixed-velocity experiments over a range of reduced frequencies, Strouhal numbers and aspect ratios relevant to bio-inspired propulsion.  The scaling laws elucidate the dominant flow physics behind the thrust production and energetics of pitching bio-propulsors, and they provide guidance for the design of bio-inspired propulsive systems. 
\end{abstract}

\section{Introduction}
Many aquatic animals efficiently propel themselves by oscillating their bodies and caudal fins in unsteady motions. During self-propelled locomotion, swimmers reach a cruising condition where there is a balance between the time-averaged thrust, generated mainly by their caudal fins, and the time-averaged drag incurred by their bodies \citep{saadat2017rules}.  By following this idea, numerous studies \citep{chopra1976large,read2003forces, dong2005wake} have investigated the performance of heaving and pitching propulsors in isolation instead of the combined body and propulsor of other studies \citep{lighthill1960note}.  For example, Garrick's theory \citep{garrick1936propulsion} extends the small amplitude theory of Theodorsen \citep{theodorsen1935general} to calculate the thrust production and power consumption of pitching and heaving two-dimensional airfoils.  Later, this theory was extended to large-amplitude oscillations \citep{scherer1968experimental} and to three-dimensions with a particular focus on the lunate tail of aquatic animals \citep{lighthill1970aquatic, chopra1976large, chopra1977hydromechanics}. \cite{cheng1984lunate}, for example, developed an asymptotic analysis for high aspect ratio surfaces oscillating at low frequencies, where the influence of wake vorticity was captured by generalizing classical lifting line theory \citep{prandtl1920theory}. \cite{karpouzian1990lunate} improved upon this asymptotic theory for lunate tails.   While these studies have provided great insights, the identification of the flow mechanisms that lead to the scaling of the thrust and power of swimmers have been elusive for cases with propulsors ranging from low to high aspect ratios, motions ranging from small to large amplitude and with nonlinearly deforming wakes.  


Several studies have built on these analytical models by using experiments to develop scaling laws for thrust and efficiency. For instance, \cite{green2008effects} characterized the thrust production of low-aspect ratio rectangular pitching panels by deriving a scaling relation that links the Strouhal number, aspect ratio and amplitude of motion. They considered an approach inspired by Prandtl's lifting line theory to account for the effects of aspect ratio on the pressure coefficient \citep{green2008effects}. In contrast, \cite{dewey2013scaling} and \cite{quinn2014unsteady} scaled the thrust forces of pitching panels with the added mass forces.  By considering both circulatory and added mass forces, \cite{floryan2017scaling} presented scaling laws for the thrust and power of heaving or pitching two-dimensional foils. Similarly, \cite{moored2018inviscid} developed scaling laws by considering circulatory forces, added mass forces, and nonlinearities that are not accounted for in classical linear theory.   

Here, we extend the scaling relations of \cite{moored2018inviscid} from two to three dimensions by varying the aspect ratio of pitching propulsors. Our new scaling relations are then verified through simulations and experiments. We show that the core two-dimensional scaling relations presented in \cite{moored2018inviscid} can be modified by combining classical scalings from aero- and hydrodynamic theory of the added mass and effects of upwash/downwash, and by accounting for the elliptical shape of trailing-edge vortices.  The scaling relations developed in this study offer a guide to speed-up the design of bio-inspired vehicles, and they provide insight into the flow physics that drive thrust production, power consumption and efficient swimming.

\section{Problem Formulation}
\subsection{Idealized Three-Dimensional Swimmer}
\begin{figure}
    \centering
    \includegraphics[width=0.99\textwidth]{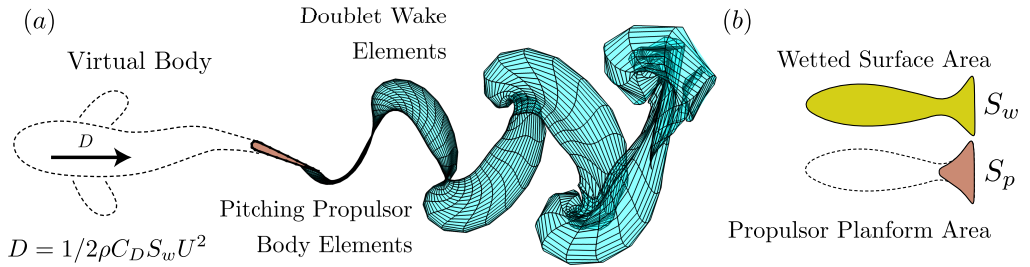}
    \caption{(a) Illustration of an idealized three-dimensional swimmer as a combination of a virtual body and a propulsor.  The doublet wake elements model vorticity shed from the trailing edge of the propulsor. (b) Representation of the wetted surface area and propulsor planform area.}
    \label{fig:VB}
\end{figure}

Self-propelled simulations are performed on an idealized swimmer that is a combination of a virtual body and a three-dimensional propulsor (Figure \ref{fig:VB}a).  The propulsor is represented by a rectangular-planform propulsor that is pitching about its leading edge.
The virtual body is not present in the computational domain, but its presence is represented as a drag force, $D$, acting on the propulsor. To determine the drag force, we used a classic high Reynolds number drag law, where drag is proportional to the square of the swimming speed $U$:
\begin{equation}
\label{e:drag}
    D = 1/2\, \rho C_D S_w U^2,
\end{equation}
where $\rho$ is the fluid density, $C_D$ is the drag coefficient and, $S_w$ represents the total wetted surface area of a swimmer (Figure \ref{fig:VB}b).

\subsection{Input Parameters}
In self-propelled swimming the time-averaged thrust and drag of a swimmer are balanced when the swimmer is at a cycle-averaged steady-state.  The drag coefficient of the body and the wetted area to propulsor planform area ratio, $S_{wp} = S_w/S_p$, both affect how thrust and drag are balanced on a swimmer and their combination is represented by the Lighthill number,
\begin{equation}
\label{e:li}
    Li = C_D S_{wp}.
\end{equation}
The Lighthill number represents the propulsor loading during self-propelled swimming and is analogous to the wing loading of birds and aircraft. When $Li$ is high there is high propulsor loading and \textit{vice versa}.  Given constant kinematics and propulsor geometries, high $Li$ swimmers will swim slower than low $Li$ swimmers.  In the current study, the Lighthill number is varied from $0.05$ to $0.15$, which covers a range typical of animal locomotion \citep{eloy2013best}.  

The non-dimensional mass of the swimmer, defined as body mass divided by added mass, $m^* \equiv  m/ \rho S_p c$, was chosen as 1.  \cite{moored2018inviscid} previously showed that the self-propelled performance of a swimmer was found to be nearly independent of its non-dimensional mass as long as $m^* \geq 1$.  By using the lower bound of this range, the simulations then reach their cycle-averaged steady-state solution with the smallest amount of simulation time.
The simulated propulsor has a chord length of $c = 0.1$ m and a NACA 0012 cross-section in accordance with previous work \cite[]{moored2018inviscid}. Its planform area is defined as $S_p = sc$, where $s$ is the span length of the propulsor.  The aspect ratio for the rectangular propulsor is defined as $\AR = s/c$, which varies from $1$ to $1000$ in the current study, where the highest aspect ratio represents an effectively two-dimensional propulsor. 

The propulsor's kinematic motion is characterized as sinusoidal pitching about the leading edge where the pitching angle is described by $\theta (t) = \theta_0\, \sin (2 \pi ft)$, where $f$ is pitching frequency, $\theta_0$ is pitching amplitude, and $t$ is time.  The pitching amplitude co-varies with peak-to-peak trailing-edge amplitude $A$, that is, $\theta_0 = \sin^{-1} (A/2c)$.  Here, we will specify peak-to-peak amplitude as a non-dimensional amplitude-to-chord ratio, 
\begin{equation}
\label{e:a_c}
    A^* = A/c.
\end{equation}

All of the input parameters used in the current study are reported in Table \ref{tab:InputParameters}.  The frequency, amplitude, and aspect ratio ranges are chosen to produce a dataset that covers the Strouhal number, reduced frequency and aspect ratio ranges that are typical of biological and bio-inspired propulsion \cite[]{saadat2017rules,sambilay1990interrelationships}.
\begin{table}
 \begin{center}
  \begin{tabular}{lccccccccc}
    \textbf{Computational Input Parameters:}\\
    $\AR$  & 1 & 1.5 & 2 & 4 & 6 & 8 & 10 & 12 & 1000 \\
    $A^*$  & 0.2 & 0.3 & 0.4 & 0.5 & 0.6 & --- & --- & --- & ---  \\
    $f$ Hz     & 1 & 2 & --- & --- & --- & --- & --- & --- & --- \\
    $m^*$ & 1 & --- & --- & --- & --- & --- & --- & --- & --- \\
    $Li$ & 0.05 & 0.1 & 0.15 & --- & --- & --- & --- & --- & --- \\
    \hline
    \textbf{Experimental Input Parameters:}\\
    $\AR$  & 1 & 1.5 & 2 & 1000 & --- & --- & --- & --- & --- \\
    $A^*$  & 0.2 & 0.3 & 0.4 & 0.5 & --- & --- & --- & --- & ---  \\
    $f$ Hz     & 0.25 & 0.5 & 0.75 & 1 & 1.25 & 1.5 & 1.75 & 2 & --- \\
    $Re$     & 30000 & --- & --- & --- & --- & --- & --- & --- & --- \\
    $U_{\infty}$ [m/s]    & 0.15 & --- & --- & --- & --- & --- & --- & --- & --- \\
  \end{tabular}
  \caption{Input parameters used in the present study.}{\label{tab:InputParameters} }
 \end{center}
\end{table}

\subsection{Output Parameters}
To examine bulk swimming performance, we time-averaged output parameters over an oscillation cycle, as indicated with an overline ($\overline{\cdot})$. All mean quantities are taken after a swimmer has reached steady state swimming, defined as the time when the net thrust coefficient is $C_{T,\text{net}} \leq 10^{-5}$, where $C_{T,\text{net}} = (\overline{T} - \overline{D}) / (1/2\, \rho S_p \overline{U}^2)$ and $T$ is the thrust force, calculated by integrating of the pressure forces projected in the $-x$ direction.  Once the mean swimming speed is determined, the reduced frequency and the Strouhal number are defined as, 
\begin{equation}
    k = \frac{fc}{\overline{U}} \;\;\;\;\;\;\;\;\; St = \frac{fA}{\overline{U}}.
\end{equation}
In self-propelled swimming, these two variables become outputs since the swimming speed is unknown \textit{a priori}.  Furthermore, the time-averaged thrust and power coefficients non-dimensionalized by the added mass forces and added mass power from small amplitude theory \citep{garrick1936propulsion} are defined as,
\begin{equation}
\label{e:GPandGT}
    C_T = \frac{\overline{T}}{\rho S_p f^2 A^2} \;\;\;\;\;\;\;\;\; C_P = \frac{\overline{P}}{\rho S_p f^2 A^2 \overline{U}}.
\end{equation}
Here, the power is calculated as the negative inner product of the force vector and velocity vector of each boundary element.  The mean thrust and power may also be non-dimensionalized by the dynamic pressure:
\begin{equation}
\label{e:DPandDT}
    C_{T,\text{dyn}} = \frac{\overline{T}}{1/2\, \rho S_p \overline{U}^2} \;\;\;\;\;\;\;\;\; C_{P,\text{dyn}} = \frac{\overline{P}}{1/2\, \rho S_p \overline{U}^3}.
\end{equation}
The two normalizations are related by simple transformations: $C_{T,\text{dyn}} =  C_T\, (2 St^2)$ and  $C_{P,\text{dyn}} = C_P \, (2 St^2)$.
\section{Methods} \label{sec:methods}
\subsection{Numerical Methods}
To model the forces acting on self-propelled pitching propulsors, we used an unsteady three-dimensional boundary element method.  We assume potential flow, that is, irrotational, incompressible and inviscid flow governed by Laplace's equation.  There is a general solution to the governing equation, so the problem is reduced to a finding a distribution of doublet and source elements on the propulsor's surface and wake that satisfy the boundary conditions.  An internal Dirichlet boundary condition is imposed in order to enforce the no-flux condition on the surface of the propulsor at each time step. The far-field boundary condition, which is that flow perturbations must decay with distance from the propulsor, is implicitly satisfied by the elementary solutions of the doublet and source elements.  The propulsor and wake surface are discretized by a finite number of quadrilateral boundary elements.  Each element on the body surface has an associated collocation point located at the element's center, just inside the body where the Dirichlet condition is enforced.  An explicit Kutta condition is enforced at the trailing-edge, and at each time step a wake doublet element is shed with a strength that satisfies Kelvin's circulation theorem.  The wake elements are advected with the local velocity field by applying the desingularized Biot-Savart Law \citep{krasny1986desingularization} leading to wake deformation and roll up.  The tangential perturbation velocity over the body is found by a local differentiation of the perturbation potential. The unsteady Bernoulli equation is then used to calculate the pressure field acting on the body. Finally, the self-propelled body dynamics are calculated when the streamwise translational degree of freedom is unconstrained. The body velocity and position are determined at the current time step through forward differencing and the trapezoidal rule, respectively. 
\begin{equation}
\label{e:SelfPropelled2}
    U_{0}^{n+1} = U_{0}^{n} + \frac{F_x^n}{M}\Delta t,
\end{equation}
\begin{equation}
\label{e:SelfPropelled1}
    x_{b}^{n+1} = x_{b}^{n} + \frac{1}{2} (U_{0}^{n+1} + U_{0}^{n})\Delta t,
\end{equation}
\noindent where $F_x^n$ is the net force acting on the foil in the streamwise direction at the $n^{th}$ timestep, $x_b$ is the body position of the foil and $\Delta t$ is the time step.  Validations of the current self-propelled boundary element method can be found in Appendix \ref{appA} for time-averaged power and velocity. More details and validations of the three-dimensional unsteady boundary element method can be found in \citep{Moored2018Bem}. Further validations and applications of the solver can be found in \citep{fish2016hydrodynamic,akoz2018unsteady,quinn2014unsteady}. 
\subsection{Experimental Methods}
\begin{figure}
    \centering
    \includegraphics[width=0.8\textwidth]{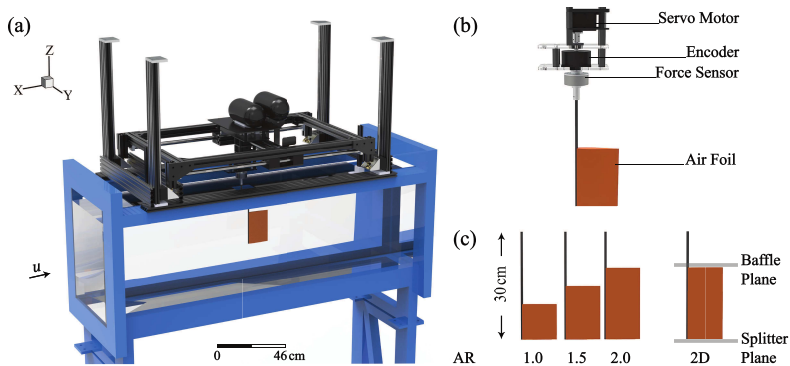}
    \caption{Experimental setup. (a) The test apparatus was mounted on the top of the water channel. (b) A servo motor actuated the airfoil using a drive rod. (c) Four different aspect ratios (\AR) were tested.}
    \label{fig:UVA_SP_FV}
\end{figure}

To complement the simulations, we measured the forces on rigid pitching airfoils suspended in a closed-loop water channel (Figure \ref{fig:UVA_SP_FV}a; Rolling Hills 1520; test section: 380 mm wide, 450 mm deep, 1520 mm long). An acrylic baffle was installed at the free surface to minimize surface waves. For all tests, the free-steam speed, $U$, was 150 mm/s with fluctuations less than 1.0\%. We tested three NACA 0012 airfoils with the same chord (100 mm) and three different spans: 100, 150, and 200 mm (aspect ratio $\AR = 1.0$, $1.5$, or $2.0$). The drive rod was made of carbon fiber and the airfoil was 3D-printed (Dimension 1200es) with ABS. We also used the 200 mm span airfoil to create a two-dimensional control case (\AR \;=\; $\infty$) by installing a horizontal splitter plate at the base of the airfoil (Figure \ref{fig:UVA_SP_FV}a). In the two-dimensional case, the gap between the airfoil tips and the baffle/splitter plate was less than 5 mm. The airfoils were actuated with harmonic pitching motions by a digital servo motor (Dynamixel MX-64). As with the simulations, the pitch angle was prescribed to be $\theta = \theta_0 \sin(2\pi f t)$. We varied the pitch frequency from 0.25 to 2.0 Hz in increments of 0.25 Hz and the non-dimensional peak-to-peak amplitude from 0.2 to 0.5 in intervals of 0.1.

We extracted the thrust and efficiency of the airfoils over a range of motions using angle and force/torque measurements. We measured pitch angle $\theta$ with an absolute encoder (US Digital A2K 4096 CPR0) and forces/torques with a 6-axis load cell (ATI MINI 40) - both of which were installed along the drive rod of the airfoil (Figure \ref{fig:UVA_SP_FV}b). The measured pitch angles and forces/torques were synchronized by a custom circuit, then transmitted (ATI Wireless F/T) to a control PC (Omen 870), where they were recorded by a custom Labview script (Labview 2017). For each trial, data were averaged over 20 pitching cycles, with 10 cycles added on either end to provide a warm-up and cool-down period. Each trial was conducted 5 times. The resolutions of the force/torque sensor were sufficient to resolve differences between the trials: force resolution was $\pm$0.01 N in the lateral (y) and streamwise (x) directions and $\pm$0.02 N in the vertical (z) direction, and torque resolution was $\pm$0.25 N$\cdot$mm about the x and y-axis and z-axis. To measure the force transmitted from the airfoil to the water, we subtracted forces measured in air (channel drained) from forces measured in water (channel filled) for all trials. We transformed the resulting forces from force sensor coordinates into water channel coordinates to get net-thrust ($T$) and lift ($L$). Subtracting forces in air produced a small effect that was most pronounced at large frequencies: when $f = 2$ Hz and $\AR = 2$, the procedure resulted in $\leq 7\%$ decreases in $T$ and $\leq 2\%$ decreases in L. The power transmitted to the fluid by the airfoil is $ P =  f \int_{t_0}^{t_0+1/f} \tau_z \dot{\theta} dt$, where $\tau_z$ is z-axis torque and $\dot{\theta}$ is pitching velocity. Coordinate transformations, phase-averaging, and power calculations were performed with Matlab R2018a.

\subsection{Two-Dimensional Scaling Relations} \label{sec:2D_Scaling}
\cite{moored2018inviscid} introduced scaling relations for the performance of two-dimensional, self-propelled pitching hydrofoils.  They considered both the added mass and circulatory forces from classical linear theory \cite[]{garrick1936propulsion} and developed a new scaling model by introducing additional nonlinear terms that are not accounted for in linear theory.  For instance, the thrust coefficient defined in eq. (\ref{e:GPandGT}), is proposed to be proportional to the superposition of three terms,  
\begin{align} \label{e:GarrickInspThrustCoeff}
    C_T &= c_1 + c_2\, \phi_2 + c_3\, \phi_3, \nonumber \\
     \mbox{with:} \; \phi_2 &= - \left[\frac{3F}{2} + \frac{F}{\pi^2 k^2} - \frac{G}{2\pi k} - \left( F^2 + G^2 \right)\left( \frac{1}{\pi^2 k^2} + \frac{9}{4} \right)\right], \\
      \phi_3 &= -A^*,  \nonumber
\end{align} 

\noindent where $c_1$, $c_2$, and $c_3$ are constants, and $F$ and $G$ are the real and imaginary components of Theodorsen's lift deficiency function, respectively \citep{theodorsen1935general}.  The first and second terms represented by $c_1$ and $c_2\, \phi_2$ are the added mass and circulatory streamwise forces, respectively, from linear theory while the third term represented by $c_3\, \phi_3$ is not accounted for in linear theory.  The third term corresponds to the form drag induced by the effects of vortex shedding at the trailing edge and it is proportional to the time-varying projected frontal area that occurs during large-amplitude pitching oscillations.
\begin{figure}
    \centering
    \includegraphics[width=0.7\linewidth]{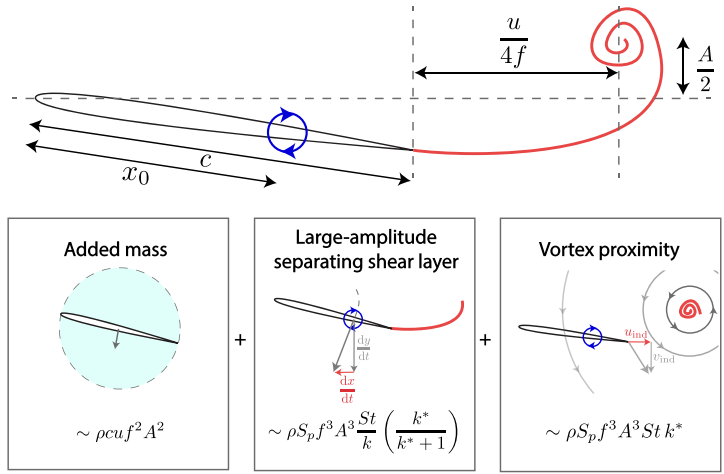}
    \caption{Schematic showing the components of the two-dimensional power scaling relation developed in \cite{moored2018inviscid}.}
    \label{fig:MooredScaling}
\end{figure}

\cite{moored2018inviscid} also proposed that the power coefficient defined in eq. (\ref{e:GPandGT}) is a linear combination of three power terms denoted graphically in Figure \ref{fig:MooredScaling} and mathematically as, 
\begin{align}
\label{e:GarrickInspPowerCoeff}
    C_P &= c_4 + c_5\, \phi_5 + c_6 \, \phi_6,  \nonumber \\
   \mbox{with:}\; \phi_5 &= \frac{St^2}{k}\left(\frac{k^*}{k^* + 1}\right), \\
   \phi_6 &= St^2 k^*, \nonumber
\end{align} 

\noindent where $c_4$, $c_5$, and $c_6$ are arbitrary constants, and $k^* = {k}/\left(1 + 4\, St^2\right)$.  The first term ($c_4$) is the added mass power from linear theory. The second term ($c_5\, \phi_5$) is a power term that is not present in linear theory and develops from the $x$-component of velocity of a pitching propulsor, which is neglected in linear theory due to a small-amplitude assumption. For large amplitude motions this velocity does not disappear, leading to an additional velocity component on the bound vorticity of the propulsor and creating an additional contribution to the generalized Kutta-Joukowski force also known as the vortex force \citep{saffman1992vortex}. The third term ($c_6\, \phi_6$) is also a power term that is absent in linear theory and develops during large-amplitude motions when the trailing-edge vortices are no longer planar as assumed in the theory.  As a result, the proximity of the trailing-edge vortices induces a streamwise velocity over the foil and an additional contribution to the vortex force.   In short, the second and third terms are described as the large-amplitude separating shear layer and vortex proximity power terms, respectively, and both terms are circulatory in nature.  For more details on the development of the two-dimensional scaling relations see \cite{moored2018inviscid}.

\section{Results}
\subsection{Propulsor Performance}
\begin{figure}
\centering
    \includegraphics[width=\linewidth]{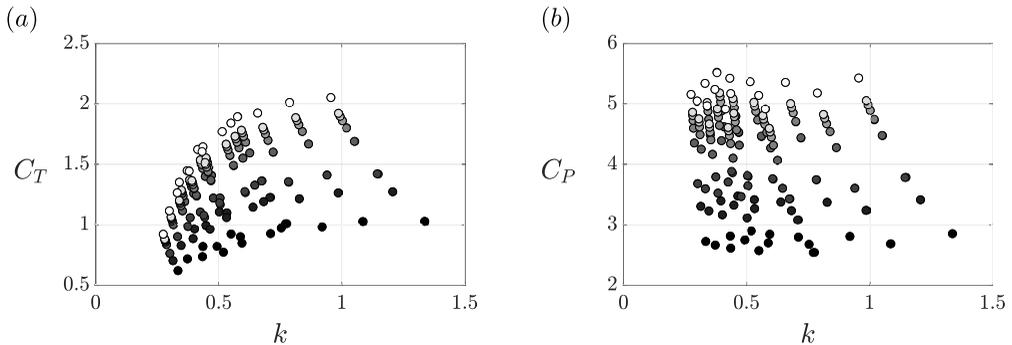}\par
    \caption{Coefficient of thrust and power as a function of reduced frequency from the self-propelled simulations. The marker colors going from black to white indicate the $\AR$ from low to high values, respectively, over the range $1 \leq \AR \leq 1000$.}
    \label{fig:coeffvsk}
\end{figure}

The combination of the computational input parameters (Table \ref{tab:InputParameters}) leads to 270 three-dimensional, self-propelled simulations with a reduced frequency range of $0.27 \leq k \leq 1.35$ and a Strouhal number range of $0.1 \leq St \leq 0.32$. From these simulations the thrust and power coefficients as defined in equation (\ref{e:GPandGT}) are calculated. Figure \ref{fig:coeffvsk} presents the thrust and power coefficients as functions of the reduced frequency. For a fixed aspect ratio, the thrust coefficient increases with the reduced frequency until an asymptotic value is reached around $k = 1$.  For a fixed reduced frequency, the thrust increases monotonically with aspect ratio, which has been observed previously \cite[]{buchholz2008wake,green2008effects}.  Like the thrust coefficient, the power coefficient generally increases with aspect ratio at a fixed reduced frequency.  For a fixed aspect ratio, the power is relatively insensitive to changes in the reduced frequency, though it shows a slight minimum around $k = 0.75$.  

It is clear that the thrust and power of pitching wings vary widely with aspect ratio.  This variation motivates corrections to the two-dimensional core scaling model developed in \cite{moored2018inviscid} that account for variations in  aspect ratio.

\subsection{Three-Dimensional Scaling Laws}
To scale the thrust and power of unsteady three-dimensional pitching propulsors, we hypothesize that the two-dimensional core scaling model will need to be modified in three ways: (1) the added mass forces must account for the added mass of a finite-span wing, (2) the circulatory forces must account for the downwash/upwash effects from the tip vortex system of the propulsor, and (3) the nonlinear wake terms must account for the elliptical topology of shedding trailing-edge vortices when calculating their induced velocity.  Here, we consider these modifications to extend the scaling relations to three-dimensional flows with propulsors of varying aspect ratio. 


\subsubsection{Added Mass and Circulatory Modifications}
In general, the added mass of an oscillating propulsor is the product of the fluid density, the planform area and a characteristic length scale of the projected area, which is commonly the chord length for two-dimensional foils. On the other hand, a characteristic added mass for arbitrary three-dimensional bodies needs to be defined by two principle dimensions in the projected area. For a rectangular propulsor, these principle dimensions are the span, $s$, and chord, $c$, of the propulsor and an empirical approximation of the added mass can be written as $M^{3D}_{add} \propto \rho s^2 c^2/(s + c)$ \cite[]{brennen1982review}.  This can be rearranged in terms of the aspect ratio noting that $\AR = s/c$ for a rectangular wing and $M_{add}^{2D} \propto \rho s c^2$ such that  $M_{add}^{3D} \propto M_{add}^{2D}\; \left[\AR/(\AR +1)\right]$.  The added mass thrust and power are both proportional to the added mass of the propulsor, so we expect them to scale with the same aspect ratio modification as the added mass itself.

A finite-span pitching wing will shed a series of vortex rings into its wake \cite[]{buchholz2008wake,green2008effects,king2018experimental}, which will lead to time-varying upwash and downwash over the wing due to trailing-edge and tip vortices.  Classical unsteady linear theory \cite[]{garrick1936propulsion} does not account for the upwash or downwash from the tip vortices, but Prandtl's finite wing theory does, at least in steady flows \cite[]{prandtl1920theory}.  We hypothesize that tip vortices would also modify the effective angle of attack of an unsteady pitching wing, and therefore that Prandtl's finite wing theory could offer a modification for the unsteady circulatory forces.  Following finite wing theory for elliptical wings, the three-dimensional lift coefficient, $C_L^{3D}$, and consequently the circulatory power, should be proportional to $C_L^{3D} \propto C_L^{2D} \left[\AR/ (\AR + 2)\right]$, where $C_L^{2D}$ is the lift coefficient from a two-dimensional foil.  Since the circulatory thrust forces of a pitching wing are the projection of this lift force in the streamwise direction, we also expect the circulatory thrust forces to scale as $C_T^{3D} \propto C_T^{2D} \left[ \AR/ (\AR + 2)\right]$.

\subsubsection{Modification of Two-Dimensional Scaling Laws}
These classical corrections from aero- and hydrodynamic theory can be applied to the circulatory and added mass terms of the two-dimensional core scaling model as follows:
\begin{equation}
\label{e:thrustcorrection}
    C_T = c_1 \left(\frac{\AR}{\AR +1}\right) - c_2\, \phi_2 \left(\frac{\AR}{\AR +2}\right)  - c_3\, \phi_3,\\
\end{equation}
\begin{equation}
\label{e:powercorrection}
    C_P = c_4 \left(\frac{\AR}{\AR +1}\right) + c_5\, \phi_5 \left(\frac{\AR}{\AR +2}\right)  + c_6\, \phi_6 \left(\frac{\AR}{\AR +2}\right).\\
\end{equation}

\noindent Here, the $c_1$ and $c_4$ terms represent added mass forces, so they were modified with the added mass correction ($\left[\AR/(\AR +1)\right]$).  In contrast, the $c_2$, $c_5$ and $c_6$ terms represent circulatory forces, so they were modified with the circulatory correction ($\left[\AR/(\AR +2)\right]$).  The $c_3$ term represents form drag and is therefore unmodified; it represents neither circulatory nor added mass forces.  Dividing by the added mass correction reveals a more compact form of the three-dimensional scaling:
\begin{align}
\label{e:thrustcorrection_ct}
    C_T^* &= c_1 - c_2\, \phi_2^* - c_3\, \phi_3^*, \\
    \mbox{where:} \; C_T^* &=  C_T \left(\frac{\AR + 1}{\AR}\right), \; \phi_2^* = \phi_2 \left(\frac{\AR + 1}{\AR + 2}\right),\; \phi_3^* = \phi_3 \left(\frac{\AR+1}{\AR}\right), \nonumber
\end{align}
\begin{align}
\label{e:powercorrection_cp}
    C_P^* &= c_4 + c_5\, \phi_5' + c_6\, \phi_6', \\
    \mbox{where:} \; C_P^* &=  C_P \left(\frac{\AR + 1}{\AR}\right),\; \phi_5' = \phi_5 \left(\frac{\AR +1 }{\AR + 2}\right), \; \phi_6' = \phi_6 \left(\frac{\AR +1 }{\AR + 2}\right). \nonumber
\end{align}

\noindent The adjusted scalings now model the three-dimensional propulsor added mass and the downwash and upwash imposed by the wake, but they do not factor in the topology of the shedding vortex system. To account for this three-dimensionality, we propose a further refinement based on the known elliptical shape of vortex rings in the wake, which we introduce in the next section.

\subsubsection{Elliptical Vortex Ring Modifications}
\begin{figure}
    \centering
    \includegraphics[width=0.9\linewidth]{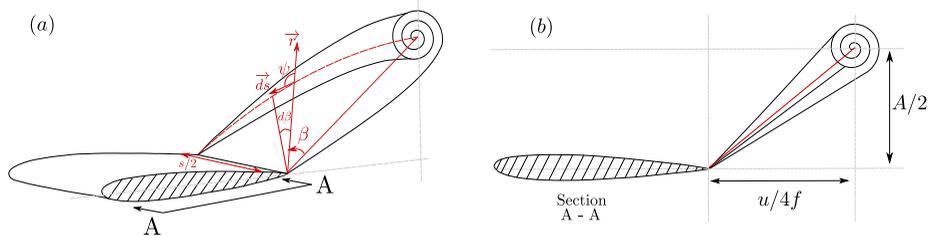}
    \caption{Sectional cut A-A is at the symmetry plane of the wing.  (a) Cross-sectional view of elliptical vortex ring shedding behind a rectangular propulsor. (b) Side view of elliptical vortex ring that shows the distance from the vortex core to the trailing edge at the mid-span of the propulsor.}
    \label{fig:ellipse}
\end{figure}
The $c_5$ and $c_6$ terms in the power scaling relation can be further modified to fully account for three-dimensionality.  Referring to the development of these terms in \cite{moored2018inviscid}, they rely on balancing the cross-stream component of the velocity induced at the trailing edge by a shedding trailing-edge vortex and the cross-stream component of the velocity induced by the bound vortex with circulation $\Gamma_b$.  This enforces the Kutta condition at the trailing edge and determines a scaling for the additional bound circulation, $\Gamma_1 = \Gamma_b - \Gamma_0$, where $\Gamma_0$ is the bound circulation from the quasi-steady motion of the wing alone, while the additional bound circulation is the bound circulation induced by the influence of the wake.  Moreover, the additional bound circulation is important to the scaling of both the $c_5$ and $c_6$ terms.  The two-dimensional scaling model assumes that the shedding trailing-edge vortex is two-dimensional, that is, it is a line vortex that extends to infinity.  In a three-dimensional flow, the shedding vortex is essentially half of a full vortex ring and is elliptical in shape (Figure \ref{fig:ellipse}), where the major axis of the ellipse is the span length of the propulsor.  This difference in the topology of the shedding trailing-edge vortex between two- and three-dimensions alters the magnitude of the induced velocity at the trailing-edge.  

By calculating the velocity induced at the trailing-edge of the propulsor mid-span by the shedding half-ellipse trailing-edge vortex (see appendix \ref{appA} for details), a new scaling relation for the additional bound circulation is
\begin{equation}
    \Gamma_1 \propto c^2  \dot{\theta} \Bigg(\frac{\gamma k^*}{1 + \gamma k^*}\Bigg) \; \; \mbox{where:} \; \; \gamma = \frac{1}{2} \left[E(m_2) + \frac{E(m_1)}{\AR \sqrt{4 k k^*}}\right].
\end{equation}

\noindent Here, $m_1$ and $m_2$ are elliptic moduli where $m_1 = \sqrt{1 - 4\AR^2 k k^*}$ and $m_2 = \sqrt{1 - 1/(4\AR^2 k k^*)}$, respectively, $E$ is the complete elliptic integral of the second kind, and $\dot{\theta}$ is the pitching rate of the propulsor.  Additionally, the vortex proximity term not only uses the additional bound circulation, but also the streamwise component of the velocity induced at the trailing edge by a shedding trailing-edge vortex.  For a shedding half-ellipse vortex, this induced velocity scales as $u_{\text{ind}} \propto c^2\dot{\theta} f St \gamma / [U (1 + 4\, St^2)]$. Consequently, the modified power scaling for the separating shear layer term, $c_5\, \phi_5^{*}$, and the vortex proximity term, $c_6\, \phi_6^{*}$, will be
\begin{equation}
    \phi_5^{*} = \frac{St^2}{k}\left(\frac{\gamma k^*}{1 + \gamma k^*}\right)\left(\frac{\AR+1}{\AR +2}\right) \;\;\;\; \mbox{and} \;\;\;\;\phi_6^{*} = St^2 k^*\gamma \left(\frac{\AR+1}{\AR +2}\right).
\end{equation}

\noindent Now, the full three-dimensional power scaling relation becomes
\begin{align}
\label{e:powercorrection_cp2}
    C_P^* &= c_4 + c_5\, \phi_5^* + c_6\, \phi_6^*
\end{align}
\noindent in its compact form.  

The scaling relations can also be written in terms of the thrust and power coefficients normalized by dynamic pressure as
\begin{align}
    C_{T,\text{dyn}}^* &= \left(c_1 + c_2\, \phi_2^* + c_3\, \phi_3^*\right) 2 St^2, \label{e:thrust_dyn} \\
    C_{P,\text{dyn}}^* &= \left(c_4 + c_5\, \phi_5^* + c_6\, \phi_6^*\right) 2 St^2, \label{e:power_dyn}
\end{align}

\noindent where $C_{T,\text{dyn}}^* = C_{T,\text{dyn}} \left[\left( \AR +1\right)/ \AR \right]$ and $C_{P,\text{dyn}}^* = C_{P,\text{dyn}} \left[\left( \AR +1\right)/ \AR \right]$, respectively. Note that the three-dimensional modifications applied to the two-dimensional core scaling model did not introduce any new terms. A summary of the scaling relations is provide in table \ref{tab:ScalingRelations}.

\subsubsection{Three-Dimensional Scaling Results}
\begin{figure}
    \centering
    \includegraphics[width=\linewidth]{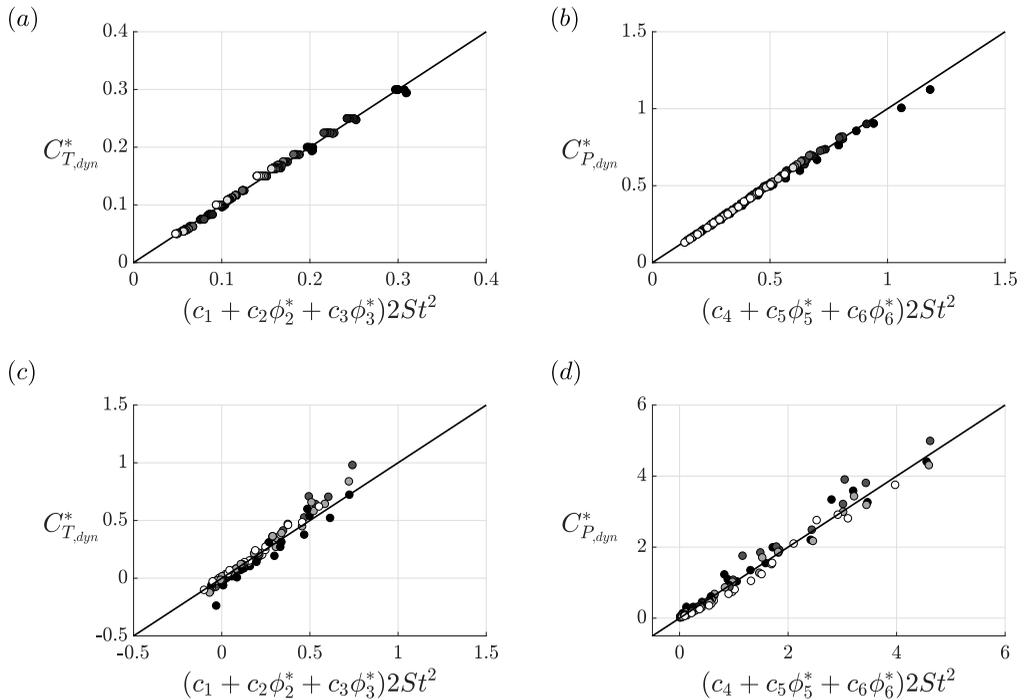}\par
    \caption{Dynamic pressure-based  thrust and power coefficients plotted as functions of their proposed scaling relations. (a,b) correspond to numerical data with the coefficients $c_1=2.83, \; c_2=3.214,\; c_3=0.5904,\; c_4=5.033,\; c_5=17.34,$ and $c_6=6.645$. (c,d) correspond to experimental data with the coefficients $c_1=3.908, \;c_2=10.9,\; c_3=0.9746,\; c_4 = 7.024\; c_5 =-64.31,$ and $c_6 = 75.97$. }
    \label{fig:3DScalingfor3D2}
\end{figure}

Figure \ref{fig:3DScalingfor3D2} presents numerical (a,b) and experimental (c,d) data for the modified dynamic pressure-based thrust and power coefficient.  Both coefficients are graphed against the scaling relations proposed in equations (\ref{e:thrust_dyn}) and (\ref{e:power_dyn}).   Note that the experimental measurements are acquired at fixed-velocity conditions and they have a reduced frequency range of $0.16 \leq k \leq 1.33$ and a Strouhal number range of $0.03 \leq St \leq 0.5$. When either the numerical or experimental data are graphed against the scaling relations, the data can be seen to collapse well to a line of slope one for both the thrust and power.  In fact, the numerical data are within $\pm 5 $\% of the scaling predictions while the experimental data are within $\pm 24$\% of the scaling predictions based on their deviation from the reference lines.  As expected the experimental data collapse is not as good as the numerical collapse, presumably due to viscous effects that are not accounted for in the numerical solutions nor the scaling relations.  Regardless, the collapse of the data to a line of slope one confirms that the newly proposed three-dimensional scaling relations capture the dominant flow physics for self-propelled or fixed-velocity pitching wings across a wide range of $k$, $St$, and $\AR$.  Moreover, an alternate geometric approach to assessing the collapse of data to three-dimensional planes can be used to show that the collapse of data is \textit{independent}
of the values of the constants $c_1$ -- $c_6$ (see Appendix \ref{appC}).

We can determine the importance of each scaling modification by considering their effects in isolation.  If only the three-dimensional added mass correction is used in the scaling relations,  the numerical data is within $\pm 40$\% and $\pm 25$\% of the thrust and power predictions, respectively, while the experimental data is within $\pm 50$\% and $\pm 45$\% of the thrust and power predictions, respectively.  If the added mass and circulatory corrections are used, the numerical data is within $\pm 5$\% and $\pm 15$\% for the thrust and power predictions, respectively, while the experimental data is within $\pm 35$\% of both scaling predictions.  Finally, when all three corrections are used, then the best agreement is recovered where the numerical data is within $\pm 5 $\% of both predictions while the experimental data is within $\pm 24 $\% of both predictions, as stated previously.  This deeper analysis shows that by applying only the added mass and circulatory corrections, a majority of the relevant flow physics can be captured, but further refinement is possible by considering the elliptical topology of the shedding trailing-edge vortices.
 \begin{table}
  \begin{center}
  \begin{tabular}{lc}
     $\overline{T} = $  & $ \rho C_T^* S_p f^2 A^2 \left[\AR /\left(\AR + 1\right)\right]$ \\
     $\overline{P} = $  & $\rho C_P^* S_p f^2 A^2 \overline{U}\left[\AR /\left(\AR + 1\right)\right]$ \\
     $C_T^* =  $  & $c_1 + c_2 \phi_2^* + c_3 \phi_3^*$ \\
     $C_P^* = $ &  $c_4 + c_5 \phi_5^* + c_6 \phi_6^*$\\
     & \\
     $\phi_2^* = $ &  $- \left[3F/2 + F/\pi^2 k^2 - G/2\pi k - \left( F^2 + G^2 \right)\left( 1/\pi^2 k^2 + 9/4 \right)\right] \left(\AR + 1\right)/ \left(\AR + 2 \right)$ \\
     $\phi_3^* = $ &  $- A^* \left[\left(\AR + 1\right)/ \AR \right]$ \\
     $\phi_5^* = $ &  $\left(St^2/k\right) \left[\gamma k^*/ \left(\gamma k^* +1\right)\right] \left(\AR + 1\right)/ \left(\AR + 2 \right)$ \\
     $\phi_6^* = $ &  $St^2 k^* \gamma \left(\AR + 1\right)/ \left(\AR + 2 \right)$ \\
     & \\
     $\gamma = $ & $ 1/2 \left[E(m_2) + E(m_1)/\left(\AR \sqrt{4 k k^*} \right) \right]$\\
     $m_1 =$ & $ \sqrt{1 - 4\AR^2 k k^*}$ \\
     $m_2 =$ & $\sqrt{1 - 1/(4\AR^2 k k^*)}$ \\
     $k^* = $ & $ k /( 1 + 4 St^2)$\\
     & \\
     
     coefficients:  & $c_1 = 3.91$ $c_2 = 10.90$ $c_3 = 0.97$ $c_4 = 7.02$ $c_5 =-64.31$ $c_6 = 75.97$\\
   \end{tabular}
   \caption{Summary of scaling relations with coefficients that are experimentally determined.}{\label{tab:ScalingRelations}}
  \end{center}
 \end{table}
\section{Conclusion}
In this work novel scaling laws of the thrust production and power consumption of pitching bio-propulsors are developed by extending the two-dimensional core scaling model presented in \cite{moored2018inviscid} to account for three-dimensionality in the form of aspect ratio variations.  This is accomplished by considering the added mass of a finite-span propulsor, the upwash/downwash effects on a propulsor from its trailing vortex system, and the influence of elliptical vortex rings shed at the trailing edge.  Both self-propelled numerical simulations and  fixed-velocity experimental measurements confirm that the proposed scalings can be used to predict thrust and power.  The established scaling relationships elucidate the dominant flow physics behind the force production and energetics of pitching bio-propulsors and can be used to accelerate the design of bio-inspired devices.  

\section{Acknowledgements}
This work was support by the Office of Naval Research under Program Director Dr. Robert Brizzolara on MURI grant number N00014-08-1-0642 as well as by the National Science Foundation under Program Director Dr. Ronald Joslin in Fluid Dynamics within CBET on NSF award number 1653181.

\appendix
\section{Validation}\label{appA} 

\subsection{Numerical Validation}
\begin{figure}
    \centering
    \includegraphics[width=1\linewidth]{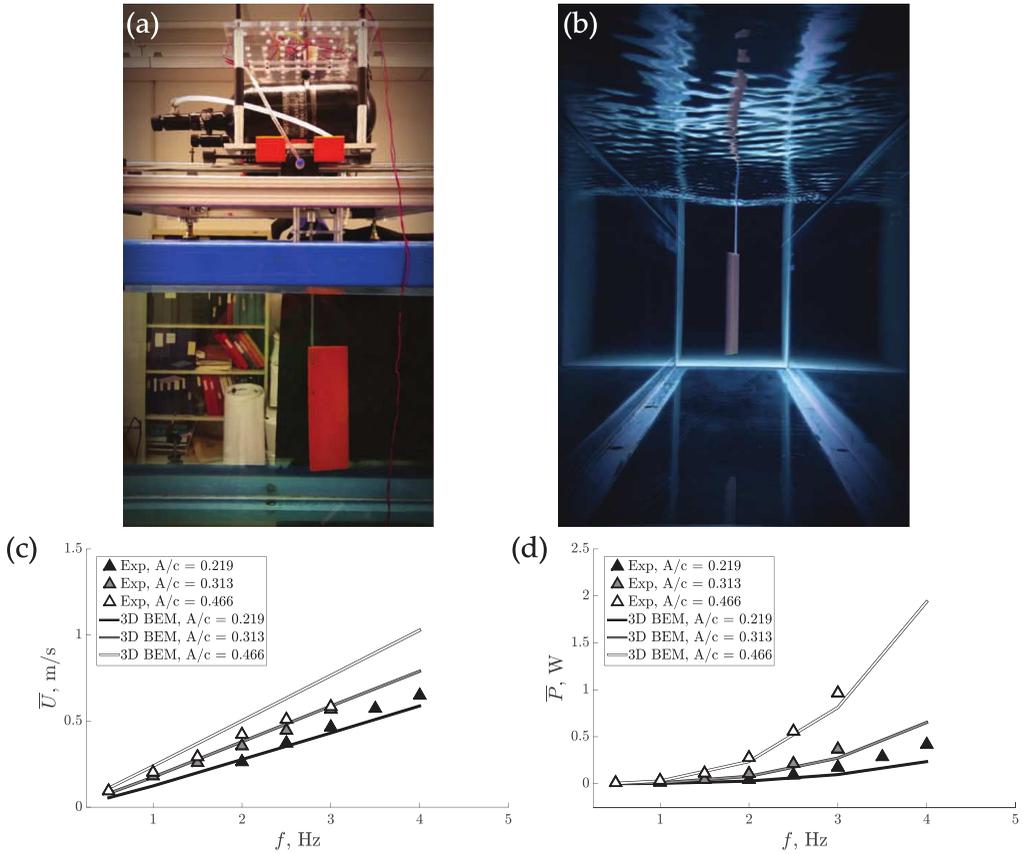}\par
    \caption{(a,b) Self-propelled experimental apparatus.  Time-averaged (c) swimming speed and (d) power as functions of pitching frequency. Numerical boundary element method solutions are denoted with solid lines while experimental measurements are denoted by triangle markers. The line and marker color, changing from black to white, corresponds to different amplitudes of motion going from the lowest to highest values, respectively.}
    \label{fig:UVAComp}
\end{figure}

In order to validate the accuracy of the self-propelled boundary element method simulations, self-propelled experiments were performed in the same closed-loop water channel as the fixed-velocity experiments (Figure \ref{fig:UVA_SP_FV}a; Rolling Hills 1520; test section: 380 mm wide, 450 mm deep, 1520 mm long).  Figure \ref{fig:UVAComp}a and \ref{fig:UVAComp}b, show the experimental apparatus, which consists of a NACA 0012 rectangular wing with an aspect ratio of two submerged in the center of the tunnel and pitched sinusoidally about its leading edge by a digital servo motor (Dynamixel MX-64).  The experiments were performed over a range of frequencies from $0.5 \leq f \leq 4$ Hz and with three non-dimensional amplitudes of $A^* = 0.219,\; 0.313, \; \mbox{and} \; 0.466$. The wing and actuation mechanism are mounted onto frictionless air bushings (Newway air bearings) that float on $0.75$" stainless steel rails oriented in the streamwise direction.  The velocity of the water tunnel is tuned until the wing is neither moving up- or downstream over several flapping cycles.  The tunnel velocity then represents the self-propelled swimming speed for a given set of kinematic parameters.  The pitching moment and the time-varying pitch angle were measured directly from the internal voltage and current sensors (used with a calibration curve to determine torque) as well as the position sensors of the Dynamixel MX-64 servo motor.

Numerically, a rectangular wing of $\AR = 2$ with a  NACA 0012 profile and pitching about its leading edge was simulated in self-propelled swimming as described in Section \ref{sec:methods}.  One difference between this validation and the simulations for the main portion of the current study is that a boundary layer solver is used to calculate the drag on the wing instead of an imposed drag force from a virtual body.  The viscous boundary layer solver uses the outer potential flow to calculate the skin friction drag using a \vK momentum integral approach.  The coupled boundary layer solver is extensively detailed and validated in previous work \citep{Moored2018Bem}.  The drag associated with an actuating rod is estimated by using the drag coefficient-Reynolds number relationship for cylinders in uniform flow to impose an additional drag force beyond the skin friction of the wing in the simulations \citep{munsonh}. The top and the bottom surfaces of the propulsor are discretized into 20 spanwise and 40 chordwise elements for a total of 3200 body elements. The computation is discretized into 50 timesteps per oscillation cycle and run for a total of 20 cycles.  All of the doublet wake elements are lumped into a set of elements once they have advected far enough downstream to change the forces by less than 1\% in order to restrict the growth of the problem size in time.

Figure \ref{fig:UVAComp}c and \ref{fig:UVAComp}d presents the time-averaged power as a function of the frequency.  The colors of the lines and markers represent the amplitude of motion with the smallest to the largest amplitudes mapped from black to white, respectively.  The solid lines represent the numerical solutions, while the triangle markers represent the experiments.  As the frequency of motion increases the swimming speed and power consumption both increase as expected.  The simulations show excellent agreement with the experiments for the self-propelled swimming speed using the two lowest amplitudes.  At the highest amplitude of motion the simulations modestly over-predict the swimming speed.  This discrepancy is likely occurring due to leading-edge separation in the experiments, which is not modeled in the simulations and is well-known to occur for high pitch amplitudes \citep{das2016existence}.  The simulations show excellent agreement with the experiments for the power consumption over all of the amplitudes examined.  The simulations are only slightly underpredicting the power at the highest frequencies.  Overall, the experiments act as a further validation of the BEM simulations presented in this work.  
\begin{figure}
    \centering
    \includegraphics[width=1\linewidth]{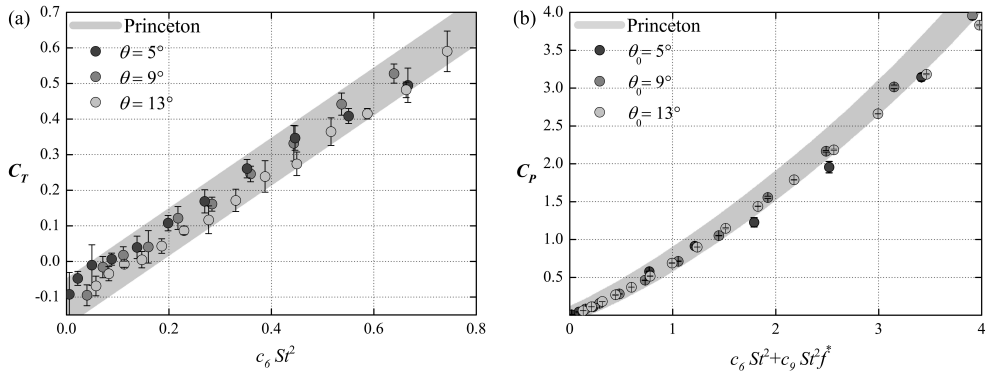}\par
    \caption{Validating thrust/power data. (a) The measured thrust coefficient agrees with published values. Thrust coefficient is defined as $C_T \equiv T/0.5 \rho a u^2 $, where $\rho$ is water density and $a$ is airfoil area (twice the span times chord). Frequency is scaled as it was in the previous study \citep{floryan2017scaling} to facilitate comparisons: the Strouhal number is defined as $St \equiv f A^* c/ u$, and the scaling constant $c_6=2.55$. Error bars show +/- one standard deviation, and the shaded band shows an envelope circumscribing the published values \citep{floryan2017scaling}. (b) The measured power coefficient also agrees with published values, particularly at lower frequencies. The reduced frequency is defined as $f^*=2 \pi f c/u$, where $c$ is chord length, and the scaling constant $c_9=4.89$.}
    \label{fig:ExpValidation}
\end{figure}

\subsection{Experimental Validation}

We validated the accuracy of our setup by comparing our force/torque data to data from a previous study in a similarly-sized closed-loop water channel at Princeton University \citep{floryan2017scaling}. As in our study, the previous study measured forces and torques on rigid pitching airfoils. Our measured time-averaged thrust and power coefficients agree very well with the published values (Figure \ref{fig:ExpValidation}a). To make a fair comparison, we matched all possible experimental conditions. First, we recreated a rigid airfoil with the same geometry: a teardrop cross section, an 80 mm chord, a maximum thickness of 8 mm, and an aspect ratio of 3.5 \citep{floryan2017scaling}. To minimize differences in vibrational noise, we used the same distance between the bottom of the force sensor and the top edge of the airfoil (2.5 cm). We used a horizontal splitter plate to match the depth of the Princeton channel (300 mm) and used the same free-stream velocity, 60 mm/s (a chord-based Reynolds number of 4780). We recreated three of the reported pitch amplitudes, $\theta_0 = 5, 9 \; \mbox{and}\; 13^o$, over a range of frequencies. Each individual trial was performed 7 times with 30 total cycles: 5 cycles for a warm-up period, 20 cycles for data acquisition, and 5 cycles for a cooling period. To increase the signal-to-noise ratio at low frequencies ($f < 0.5$ Hz), we applied a 2nd-order Butterworth filter (6$f$ cutoff frequency) to all instantaneous force and position data. The only difference between setups was that the previous study’s airfoil was a single piece of anodized aluminium, whereas ours was 3D-printed (Dimension 1200es) with ABS and fixed to a carbon fiber drive rod. To be consistent with the previous study, we did not remove inertial forces by subtracting forces in air from forces in water. Given the lower frequencies used for our validation, we do not expect this difference to significantly affect our comparison, though it may explain the slight discrepancy between the two studies in the power reported at higher frequencies (Figure \ref{fig:ExpValidation}b).

\section{Induced Velocity from a Half-Ellipse Shaped Vortex}\label{appB} 
The induced velocity at the midspan trailing edge of a wing from a half-ellipse shaped shedding vortex ring is a function of the radius of the ellipse, $r$, and the eccentric anomaly, $\beta$, (Figure \ref{fig:ellipse}). The Biot-Savart law provides a general description of the differential induced velocity field produced by a differential segment of a vortex, while the total velocity will be the integration of this influence along the length of the vortex as,
\begin{equation}\label{e:biot0}
   \mathbf{V_\textbf{ind}} = \frac{\Gamma}{4 \pi} \int_{0}^{\pi}{\frac{ \mathbf{ds} \times \textbf{r}}{|r|^3}} \;\;\; \mbox{where:} \;\;\;
    \mathbf{r}(\beta) = \frac{(s/2) \; \mathbf{r_s}}{\sqrt{(s/2)^2 \; \sin^2 \beta + r_s^2 \; \cos^2 \beta}}.
\end{equation}

\noindent Here $\Gamma$ is the vortex circulation, $\textbf{ds}$ is a vector describing the length and orientation of a differential segment of a vortex, $\textbf{r}$ is a vector from the midspan trailing edge to $\textbf{ds}$, $s$ is the span length, and $\mathbf{r_s}$ is the $\mathbf{r}$ vector at the symmetry plane of the wing, which is defined as $\mathbf{r_s} = U/(4f)\, \mathbf{i} + A/2\, \mathbf{j}$ as shown in Figure \ref{fig:ellipse}.  The cross product between $\textbf{ds}$ and $\textbf{r}$ can be written as,
\begin{equation}
    \mathbf{ds} \times \mathbf{r} = |ds||r|\sin \psi
    \label{e:biot1}
\end{equation}
where $\psi$ is the angle between the vectors $\textbf{ds}$ and $\textbf{r}$ and $\sin \psi$ will be, 
\begin{equation} \label{e:biot2}
    \sin\psi = |r| \frac{d\beta}{ds}.
\end{equation}
By substituting (\ref{e:biot1}) and (\ref{e:biot2}) into (\ref{e:biot0}) the magnitude of the induced velocity becomes \begin{equation}
\label{e:ind_e}
    \left|V_\text{ind}\right| = \frac{\Gamma}{4 \pi} \int_{0}^{\pi}{\frac{d\beta}{(s/2) \; |r_s|}\sqrt{(s/2)^2\; \sin^2 \beta + r_s^2 \; \cos^2 \beta}}
\end{equation}
Equation (\ref{e:ind_e}) is an elliptic integral of the second kind with the solution, 
\begin{equation}
    \left|V_\text{ind}\right| = \frac{\Gamma}{4 \pi} \frac{(s/2)\; E(m_2) + |r_s| \; E(m_1)}{(s/2) \; |r_s| }.
    \label{e:u_ind_phi5}
\end{equation}
Here $E$ is the complete elliptical integral of the second kind and $m_1$  and $m_2$ are the elliptic moduli defined as, 
\begin{equation}
    m_1 = \sqrt{1 - \frac{(s/2)^2}{r_s^2}} \; \; \; \mbox{and} \; \; \; m_2 = \sqrt{1 - \frac{r_s^2}{(s/2)^2}}.
    \label{e:e_modulus}
\end{equation}
Equations (\ref{e:u_ind_phi5}) and (\ref{e:e_modulus}) can be rewritten as,
\begin{equation}
    \left|V_\text{ind}\right| = \frac{\Gamma}{4 \pi} \Bigg(\frac{\AR\; c \;E(m_2) + 2\;|r_s|\; E(m_1)}{\AR \; c\; |r_s|}\Bigg),\;\;\mbox{and} \;\; |r_s| = {\sqrt{ \left(\frac{A}{2}\right)^2 + \left(\frac{U}{4 f}\right)^2}}
\end{equation}
\begin{equation}
    m_1 = \; \sqrt{1 - 4\AR^2 k k^*} \; \; \; \mbox{and} \; \; \; m_2 = \; \sqrt{1 - 1/(4\AR^2 k  k^*)}
\end{equation}
The direction of the induced velocity is mutually perpendicular to the direction of $\mathbf{r_s}$ and the spanwise direction.

\section{Alternate Geometric Assessment of Data Collapse}\label{appC} 
\begin{figure}
    \centering
    \includegraphics[width=\linewidth]{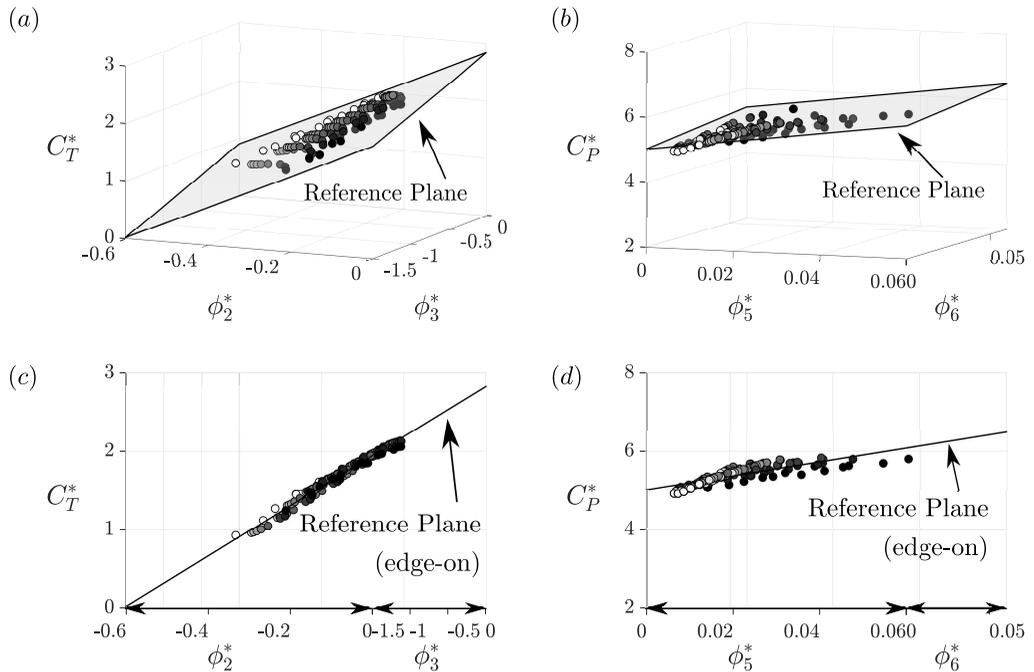}\par
    \caption{(a,b) Three-dimensional graphs of the thrust and power coefficients as functions of their scaling terms. (c,d) Three-dimensional thrust and power coefficient graphs oriented edge-on with a reference plane. The marker colors indicate the aspect ratio values varying between $1 \leq \AR \leq 1000$ with a gradient of color from black to white, respectively.}
    \label{fig:3DScalingfor3D}
\end{figure}

The thrust and power scaling relations stated in equations (\ref{e:thrustcorrection_ct}) and (\ref{e:powercorrection_cp}), respectively, represent flat planes in three-dimensions.  If the relations are accurate, then $C_T^*$ graphed as a function of $\phi_2^*$ and $\phi_3^*$ should collapse to a flat plane.  Similarly, $C_P^*$ graphed  as a function of $\phi_5^*$ and $\phi_6^*$ should also collapse to a flat plane.  Figure \ref{fig:3DScalingfor3D}a and \ref{fig:3DScalingfor3D}b present $C_T^*$ and $C_P^*$ plotted as functions of their scaling terms. By rotating the orientation of the data about the $C_T^*$ and $C_P^*$ axes such that the data is viewed ``edge-on" (Figure \ref{fig:3DScalingfor3D}c and \ref{fig:3DScalingfor3D}d), it becomes clear that there is an excellent collapse of the data to flat planes.  In fact, the scaling laws are accurate to within $\pm 5$\% of their full-scale value based on the deviation of the data from the reference planes.  The collapse occurs over wide ranges of $k$, $St$ and $\AR$, and is \textit{independent} of the values of the constants $c_1$ -- $c_6$.  Moreover, the scaling relations can be used as predictive relations once the values of the constants are determined. 

Using this geometric approach to show the collapse of experimental data is problematic.  The issue is that the uncertainties in the measurements at low frequencies and low amplitudes are amplified when the thrust and power are normalized by $f^2 A^2$, which can lead to a misinterpretation of the collapse.  Thus, assessing the collapse of experimental data with the proposed scaling relations is best accomplished when the thrust and power are normalized by dynamic pressure as presented in the main body of this study.  However, since this three-dimensional geometric approach is equivalent to the approach in the main body, it can be concluded that the collapse of the data \textit{does not} depend on the values of the constants $c_1$ -- $c_6$.


\bibliographystyle{jfm}

\end{document}